\documentclass[aps,amssymb]{revtex4}
\usepackage{graphicx}
\newcommand{\ket}[1]{|#1\rangle}
\newcommand{\bra}[1]{\langle#1|}

\begin{document}

\title{Thermodynamical Cost of Accessing Quantum Information}
\author{Koji Maruyama$^1$, \v{C}aslav Brukner$^{2}$, and Vlatko
Vedral$^{2,3}$}
\affiliation{$^1$Laboratoire d'Information Quantique and QUIC, CP 165/59, Universit\'{e} Libre de
Bruxelles, Avenue F.D. Roosevelt 50, 1050 Bruxelles, Belgium\\
$^2$Insitut f\"{u}r
Experimentalphysik, Universit\"{a}t Wien, Boltzmanngasse 5,
A-1090 Wien, Austria\\
$^3$The School of Physics and Astronomy, University of Leeds, Leeds LS2 9JT, United
Kingdom}
\date{\today}

\begin{abstract}
Thermodynamics is a macroscopic physical theory whose two very general laws are
independent of any underlying dynamical laws and structures. Nevertheless, its
generality enables us to understand a broad spectrum of phenomena in physics,
information science and biology. Does thermodynamics then imply any results in quantum
information theory? Taking accessible information in a system as an example, we show
that thermodynamics implies a weaker bound on it than the quantum mechanical one (the
Holevo bound). In other words, if any post-quantum physics should allow more information
storage it could still be under the umbrella of thermodynamics.
\end{abstract}
 \maketitle

Since ``information is physical" \cite{landauer91} the performance of information
theoretic tasks are ultimately governed by the underlying physical laws used to process
it. For example, in quantum mechanics, information that can be stored or accessed is
limited by the Holevo bound \cite{holevo73}. On the other hand, information theory is
also deeply connected with thermodynamics as most notably demonstrated by the resolution
of the long-standing Maxwell's demon paradox \cite{maxwell,demon} on the basis of
Landauer's erasure principle \cite{landauer61_bennett82}. The insight acquired from
Landauer's principle enabled the demon paradox to be extended to the quantum regime
\cite{demon,zurek84,lubkin87,lloyd97} and its link with limits on efficiency of certain
quantum information processing \cite{nielsenchuang00} has also been established
\cite{martin9901,vedral00}. The amount of heat convertible into work in reversible and
irreversible processes was considered in the context of quantum distinguishability
\cite{lyuboshitz70_levitin93} and it was shown that distinguishing non-orthogonal states
perfectly would lead to the violation of the second law of thermodynamics \cite{peres93}.
The latter suggests that altering quantum laws may have dramatic consequences on our
other theories.

Here we derive a thermodynamical bound on accessible information in quantum mechanics
from the second law of thermodynamics, which states in Kelvin's form ``There is no
thermodynamical cycle whose sole effect is the conversion of heat withdrawn from a
reservoir into mechanical work." The background of our motivation is the fact that the
generality of thermodynamical laws has led physicists to derive many, at first sight
unrelated, results, such as general relativity \cite{bekenstein73_jacobson95}, the
superposition principle in quantum mechanics \cite{lande52}, and the wave nature of
light \cite{gabor61} to name a few. In this paper, we investigate what constraint the
second law imposes on accessible information and show that thermodynamical bound is
\textit{weaker} than the Holevo bound.

Assumptions we make here are, (a) Entropy: the von Neumann entropy is equivalent to the
thermodynmaical entropy, (b) Statics and measurement: a physical state is described by a
``density" matrix, and the state after a measurement is a new state that corresponds to
the outcome (``projection postulate"), (c) Dynamics: there exist isentropic
transformations. These rules can also describe probability distributions in classical
phase space. Although we will use Dirac's ket notation for convenience, this does not
mean that we use the full machinery of the Hilbert spaces (such as the notion of inner
product) and we never use the Born trace rule for calculating probabilities.

By accessible information we mean information obtained from an arbitrary measurement on a
given system. To give a precise form of the Holevo bound let us consider two
protagonists, Alice and Bob. Suppose Alice has a classical information source preparing
symbols $i=1,...,n$ with probabilities $p_1,...,p_n$. Bob attempts to determine the
actual preparation $i$ as best he can. Thus, after Alice prepared a state $\rho_i$ with
probability $p_i$ and gives the state to Bob, who makes a general quantum measurement
(\textit{Positive Operator Valued Measure} or \textit{POVM}) with elements
${E_j}={E_1,...,E_m}$, $\sum_{j=1}^{m} E_j =\mathbf{l}$, on that state. On the basis of
the measurement result he infers Alice's preparation $i$. The Holevo bound is an upper
bound on accessible information, i.e.
\begin{equation}\label{holevo}
I(A:B)\le S(\rho)-\sum_i p_i S(\rho_i),
\end{equation}
where $I(A:B)$ is the mutual information between the set of Alice's preparations $i$ and
Bob's measurement outcomes $j$, $S(\rho)=-\text{Tr}(\rho\log_2\rho)$ is the von Neumann
entropy and $\rho=\sum_i^n p_i \rho_i$. The equality in expression (\ref{holevo}) is
achieved if all $\rho_i$ mutually commute, that is, $[\rho_i,\rho_j]=0$ for all $i,j$,
and the measurement is performed in the joint eigenbasis of $\rho_i$'s. We will refer to
this case as ``classical", as it corresponds to distinguishing between classical
probability distributions.

Let us now derive thermodynamical bound on the mutual information $I(A:B)$. To this end
we will consider a thermodynamical loop that involves a conversion of heat into work,
whose amount is equal to $kT\ln2 I(A:B)$ (throughout the paper we call $kT\ln2$ unit of
work as one bit. Here $k$ is the Boltzmann constant and $T$ is the temperature).
Examining the condition on the work balance imposed by the second law, we will have a
thermodynamical bound on $I(A:B)$.

Consider a vessel of volume $V$ filled with $N$ molecules of dilute, inert, ideal gas.
Suppose that $p_1 N$ molecules occupy the space on the left side ($L$) of the vessel,
whose volume is $p_1 V$, and each individual molecule is in the quantum state
$\ket{\psi_1}$. Similarly, $p_2 N$ molecules ($p_1+p_2=1$) are in the right side ($R$)
of volume $p_2 V$ and are all in $\ket{\psi_2}$. The two states, $\ket{\psi_1}$ and
$\ket{\psi_2}$, can be thought of as the states of an internal degree of freedom such as
spin. The two types of molecules are initially separated by a partition and the
pressures on both sides are equal.
Note that this situation differs from the encoding/decoding scenario given above in which
Bob has no access to spatial degree of freedom but can only measure internal degree of
freedom of the molecules. Even though we primarily deal with only two pure quantum states
and projective measurements with two possible outcomes, our consideration can easily be
generalized to arbitrary numbers of general states and measurement outcomes.

We can now have a thermodynamical loop formed by two different paths between the above
initial thermodynamical state to the final state. In the final state, both constituents,
$\ket{\psi_1}$ and $\ket{\psi_2}$, will be distributed uniformly over the whole volume
of the vessel. Hence, each molecule in the final state can be described by $\rho=\sum_i
p_i \ket{\psi_i}\bra{\psi_i}$, regardless of the position in the vessel. One of the paths
converts heat into work, involving measurement (and thus it is irreversible, in general),
while the other path, consisting of a quasi-static reversible process and isentropic
transformations, requires some work consumption.

The work-extracting process proceeds as follows. Suppose that we have two semipermeable
membranes, $M_1$ and $M_2$, which separate two perfectly distinguishable (orthogonal)
states $\ket{e_1}$ and $\ket{e_2}(=\ket{e_1^\perp})$. These membranes were considered by
von Neumann \cite{vonneumann} and Peres \cite{peres93} and shown to be physically
legitimate. The membrane $M_1$ acts as a completely opaque wall to molecules in
$\ket{e_1}$, but it is transparent to molecules in $\ket{e_2}$. Similarly, $M_2$ is
opaque to molecules in $\ket{e_2}$ and transparent to $\ket{e_1}$. Thus, for example, a
state $\ket{\psi_i}$ is reflected by $M_1$ to become $\ket{e_1}$ with (conditional)
probability $p(e_1|\psi_i)$ and goes through with probability $p(e_2|\psi_i)$, being
projected onto $\ket{e_2}$. This corresponds to the quantum (projective) measurement on
molecules in the basis $\{\ket{e_1},\ket{e_2}\}$, however, we do not compute these
probabilities specifically as stated above.

\begin{figure}
 \begin{center}
  \includegraphics[scale=0.5]{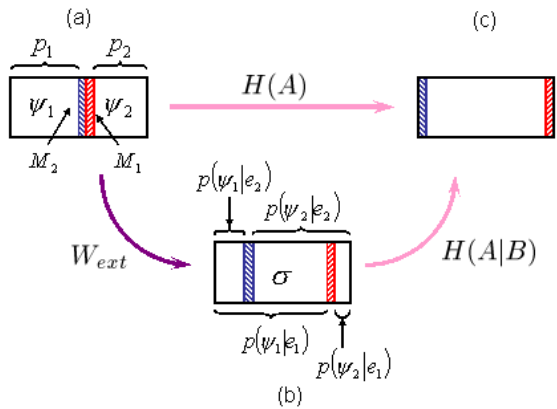}
  \caption{Equivalence between the extractable work $W_{ext}$ and the
accessible information $I(A:B)$. Suppose that each $\ket{\psi_i}$ in the initial
preparation was correlated with a state of another degree of freedom so that in the
initial state (a) there are $\ket{\psi_1}\ket{L}$ on the left of the vessel and
$\ket{\psi_2}\ket{R}$ on the right. As this auxiliary system is hypothetical, we cannot
access to this degree of freedom throughout the process discussed in the main text.
Nevertheless, if we could make use of membranes that distinguish $\ket{L}$ and $\ket{R}$,
then we can extract $H(A)$ bits of work to reach the state (c), where
$H(A)=H(p_i)=-\sum_i p_i\log_2 p_i$ is the Shannon entropy. If we use the ``proper"
membranes, $M_1$ and $M_2$, that measure $\ket{\psi_i}$ in a basis $\{\ket{e_i}\}$,
$W_{ext}$ bits of work will be extracted, and if $\ket{\psi_1}$ and $\ket{\psi_2}$ are
not perfectly distinguishable the membranes will stop before reaching the end of vessel
(as in (b)). Let us consider the gas in $\ket{e_1}$, for example. In (b), the numbers of
$\ket{e_1}$-molecules in the left and the right sides of $M_1$ are
$p(\psi_1)p(e_1|\psi_1)N=p(e_1)p(\psi_1|e_1)N$ and $p(e_1)p(\psi_2|e_1)N$, respectively,
where $p(x)$ is the proportion of $\ket{x}$ to $N$ (thus $p(\psi_i)=p_i$), and $p(x|y)$
represents the probability of finding $\ket{x}$ in $\ket{y}$. By using the same
membranes as those used in the direct path from (a) to (c) (namely, distinguishing
$\ket{e_1}\ket{L}$ and $\ket{e_1}\ket{R}$ that are separated by $M_1$), $H(A|B)$ bits of
work can be extracted in the process from (b) to (c). As these hypothetical
work-extraction processes with the auxiliary system are quasi-static and reversible, a
simple relation, $H(A)=W_{ext}+H(A|B)$, holds and this means $W_{ext}=I(A:B)$.}
 \label{accinfo}
 \end{center}
\end{figure}

\begin{figure}
 \begin{center}
  \includegraphics[scale=0.5]{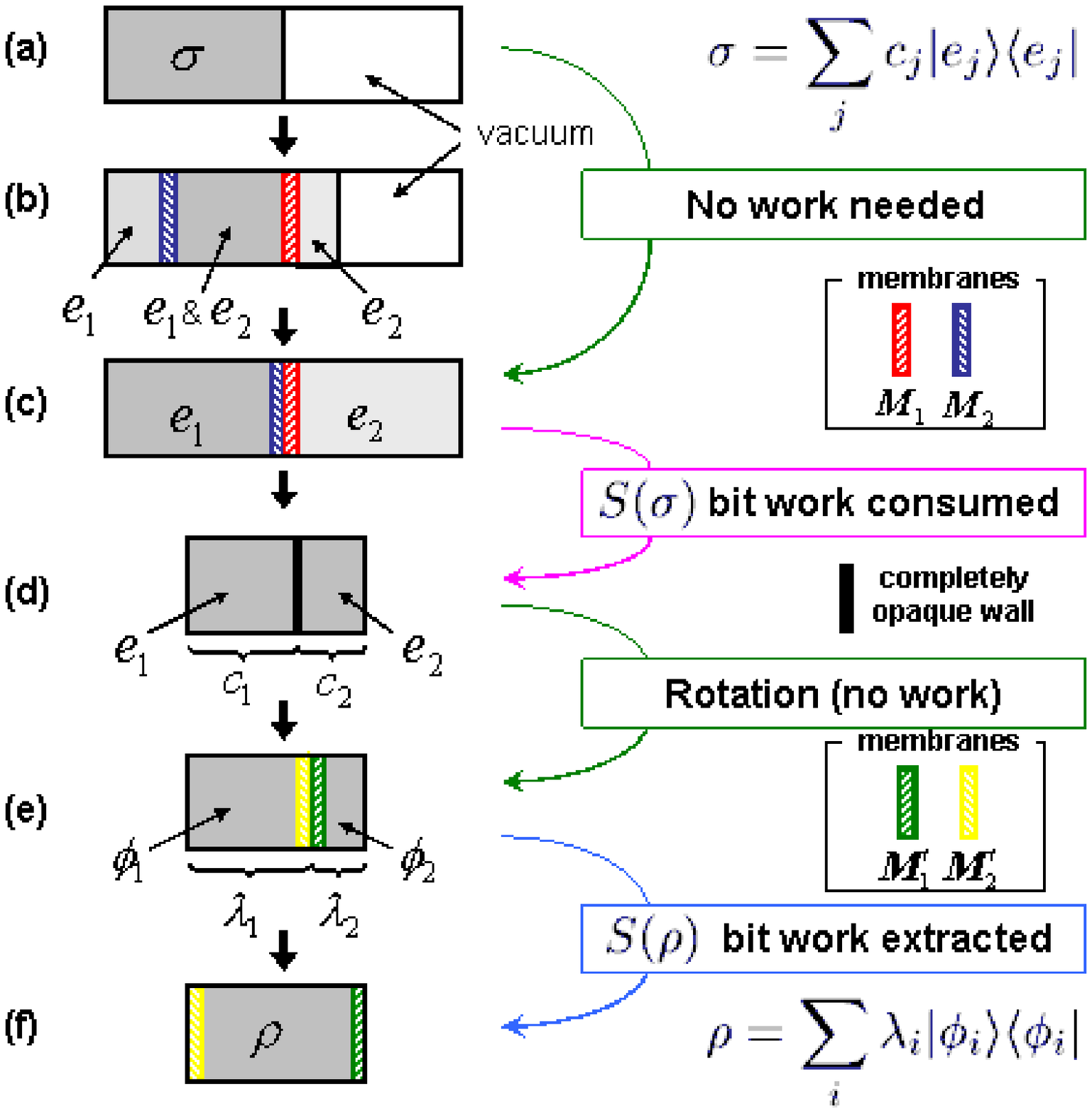}
  \caption{The thermodynamical process to transform $\sigma$ into $\rho$.
Firstly, after attaching an empty vessel of the same volume to that containing the gas
$\sigma$, the membranes $M_j$ are used to separate two orthogonal states $\ket{e_1}$ and
$\ket{e_2}$ ((a) to (c)). As the distance between the movable opaque wall and the
membrane $M_2$ is kept constant, this process entails no work consumption/extraction. As
$\sigma=\sum c_j\ket{e_j}\bra{e_j}$, compressing each $\ket{e_j}$-gas into the volume of
$c_jV$ as in (d) makes the pressures of gases equal and this compression requires
$S(\sigma)=-\sum c_j\log_2 c_j$ bits of work. Secondly, quantum states of gases are
isentropically transformed, thus without consuming work, so that the resulting state (e)
will have $\lambda_j N$ molecules in $\ket{\phi_j}$, where $\rho=\sum
\lambda_j\ket{\phi_j}\bra{\phi_j}$ is the eigendecomposition of $\rho$.
To reach (f), $S(\rho)$ bits of work can be extracted by using membranes that
distinguish $\ket{\phi_j}$. As a result, the work needed for the transformation
$\sigma\rightarrow\rho$ is $S(\sigma)-S(\rho)$ bits.}
 \label{sigma2rho}
 \end{center}
\end{figure}

We replace the partition separating $\ket{\psi_1}$ and $\ket{\psi_2}$ with the two
membranes, $M_1$ and $M_2$. Keeping its temperature constant by contact with a heat bath
of temperature $T$, each gas of molecules in $\ket{e_1}$ or $\ket{e_2}$ can expand
isothermally until the pressures of each gas component at both sides of a membrane
become equal.
The amount of the mechanical work $W_{ext}$, which can be withdrawn from the heat bath,
is equal to the accessible information $I(A:B)$, which is the amount of information Bob
can obtain about Alice's preparation by measurement on $\rho=\sum_i p_i
\ket{\psi_i}\bra{\psi_i}$ in the basis $\{e_1, e_2\}$. Figure \ref{accinfo} shows the
equivalence between $W_{ext}$ and $I(A:B)$. The same correspondence exists also if the
initial state is a collection of mixed states, such as $\{p_i,\rho_i\}$, which means that
Alice provides $\rho_i$ with probability $p_i$. The transformation from the
post-work-extraction state, which we call $\sigma$ hereafter, to the final state $\rho$
can be done by a process shown in Fig. \ref{sigma2rho} and the minimum work needed is
given by $\Delta S=S(\sigma)-S(\rho)$.

There is an alternative way to look at the process with semipermeable membranes.
Maxwell's demon, who is sitting somewhere in (or by) the box, measures the state of each
molecule in the basis $\{\ket{e_1},\ket{e_2}\}$ and memorizes all outcomes from the
measurement on all molecules. Depending upon the actual outcome he operates a membrane
$M_i$ (i.e. ``controls tiny doors" on $M_i$) so that only $\ket{e_i}$ is reflected and
the other state can go through. One may then ask whether or not some work needs to be
consumed to erase the information recorded in his memory at a certain stage to close the
thermodynamical cycle. We show that it is not necessary. Let $\ket{m}$ denote the demon's
memory for outcome $m$. After the observation and work-extraction by demon, the joint
system of the principal system $P$ and the memory $M$, $\sigma^{PM}$, can be described as
$\sigma^{PM}=\sum_m P_m \rho P_m \otimes \ket{m}\bra{m}$, where $P_m=\ket{e_m}\bra{e_m}$
are projection operators. Hence, unlike the usual discussion of erasure principle with
Szilard's engine \cite{szilard29}, erasing demon's memory in this case is a logically
reversible process \cite{landauer61_bennett82} due to a perfect correlation between $P$
and $M$. Therefore a controlled-NOT-like global (isentropic) operation between $P$ and
$M$ can reset the state of $M$ to a standard initial state without consuming work. We can
consider that isentropic transformations, in principle, involve no work transfer
\cite{vonneumann}. The difference from the memory erasure in the Szilard model is that
the degrees of freedom used for the work-extraction and measurement are, in our case,
not the same. Here the external (spatial) and the internal (spin) ones are used, while
only the external one is employed in the Szilard model.

Another path, which is reversible, from the initial state to the final state is as
follows. Let $\{\ket{\phi_1},\ket{\phi_2}\}$ be an orthogonal basis which diagonalizes
the density matrix $\rho$, such that $\rho=\sum_i p_i \ket{\psi_i}\bra{\psi_i}=\sum_k
\lambda_k \ket{\phi_k}\bra{\phi_k}$, where $\lambda_k$ are eigenvalues of $\rho$. Since
any of $\{\ket{\psi_1},\ket{\psi_2}\}$ and $\{\ket{\phi_1},\ket{\phi_2}\}$ is a pure
state, appropriate isentropic transformations can transform the initial state to a state
in which $\lambda_1 V$ of the vessel on the left is occupied with $\ket{\phi_1}$ and
$\lambda_2 V$ on the right with $\ket{\phi_2}$. By using new semipermeable membranes
$M_1^\prime$ and $M_2^\prime$, which distinguish $\ket{\phi_1}$ and $\ket{\phi_2}$
perfectly, we obtain a state $\rho$ uniformly distributed over the volume $V$, after
gaining $S(\rho)$ bits of work. As this transformation from $\{p_i,\ket{\psi_i}\}$ to
$\rho$ via $\{\lambda_i,\ket{\phi_i}\}$ can be carried out reversibly, the initial state
can be restored from $\rho$ by consuming $S(\rho)$ bits of work.

If the initial state is a combination of mixed states with corresponding weights as
$\{p_i,\rho_i\}$, the extractable work to reach $\rho=\sum_i p_i\rho_i$ becomes
$S(\rho)-\sum_i p_i S(\rho_i)$. This can be seen by considering a process
$\{p_i,\rho_i\}\stackrel{\mbox{\small{(i)}}}{\longrightarrow}\{p_i\mu_j^i,\ket{\omega_j^i}\}
\stackrel{\mbox{\small{(ii)}}}{\longrightarrow}\{\lambda_k,\ket{\phi_k}\}
\stackrel{\mbox{\small{(iii)}}}{\longrightarrow}\rho$, where
$\{\mu_j^i,\ket{\omega_j^i}\}$ and $\{\lambda_k,\ket{\phi_k}\}$ are the sets of
eigenvalues and eigenvectors of $\rho_i$ and $\rho$, respectively. The process (i) needs
$\sum_i p_i S(\rho_i)$ bits of work to be consumed, and similarly the process (iii)
provides $S(\rho)$ bits of work to us. As the process (ii) involves only isentropic
transformations, nothing needs to be written in the work account book. As a result,
$S(\rho)-\sum_i p_i S(\rho_i)$ bits of work will be extracted.

\begin{figure}
 \begin{center}
  \includegraphics[scale=0.5]{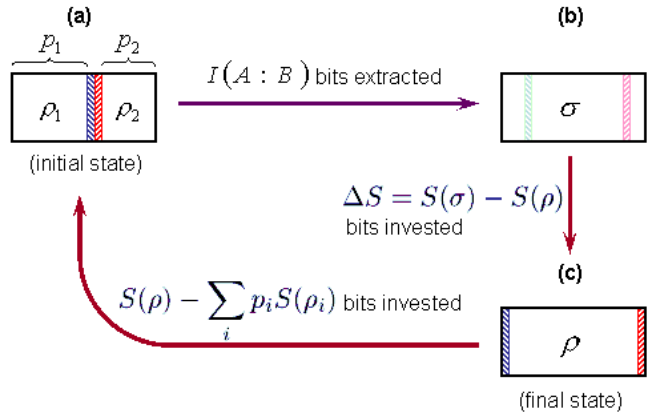}
  \caption{The thermodynamical cycle to discuss the second law. The cycle
proceeds from the initial state (a) to the final state $\rho$ (c) via the
post-work-extraction state $\sigma$ (b), and returns to the initial state with a
reversible process. The existence of a heat reservoir at temperature $T$ is assumed for
isothermal expansion/compression processes.
}
 \label{cycle}
 \end{center}
\end{figure}

Now we can discuss what the second law requires for the thermodynamical loop, which
proceeds as $\{p_i,\rho_i\} \rightarrow \sigma \rightarrow \rho \rightarrow
\{p_i,\rho_i\}$ (See Fig. \ref{cycle}). The second law states that the \textit{net} work
extractable from a heat bath cannot be positive after completing a cycle, i.e.
$W_{ext}-W_{inv}\le 0$. For the cycle described above, it can be expressed as
\begin{equation}\label{seclaw}
I(A:B)\le S(\rho)-\sum_i p_i S(\rho_i)+\Delta S,
\end{equation}
where $\Delta S=S(\sigma)-S(\rho)$. Note that $\sigma$ is identical to the resulting
state of a projective measurement on $\rho$ in the basis $\{\ket{e_1},\ket{e_2}\}$. Thus,
$\sigma=\sum_j P_j\rho P_j$ with $P_j=\ket{e_j}\bra{e_j}$ and consequently $\Delta S$ is
always non-negative \cite{vonneumann}. The inequality (\ref{seclaw}) holds even if the
measurement by membranes was a generalized (POVM) measurement. This is because any POVM
measurement on a principal system $P$ can be realized by introducing an auxiliary system
(environment) $E$ and performing a projective measurement on $E$ after letting $P$ and
$E$ interact with each other under an appropriate global unitary (thus isentropic)
evolution \cite{nielsenchuang00}. Even in this case, $\Delta S$ can be easily shown to
be non-negative by using the fact that appending a pure state to the principal system
does not change the entropy.

The form of Eq. (\ref{seclaw}) is identical with that of Eq. (\ref{holevo}) for the
Holevo bound, except for an extra non-negative term, $\Delta S$. The existence of $\Delta
S$ is essential in the cycle, where $I(A:B)$ bits of work is extracted, since the
returning path (from (b) to (c) in Fig. \ref{cycle}) is reversible, thus optimal. This
illustrates that there is a difference between the bound imposed by quantum mechanics
(the Holevo bound) and the one imposed by the second law of thermodynamics. Namely, there
is a region in which we could violate quantum mechanics while complying with the
thermodynamical law. In the classical limit, the measurement is performed in the joint
eigenbasis of mutually commuting $\rho_i$'s, consequently $\Delta S=0$, and in addition
the Holevo bound is saturated: $I(A:B)= S(\rho)-\sum_i p_i S(\rho_i)$. Thus, the
classical limit and the thermodynamical treatment give the same bound.

The same saturation occurs when an appropriate collective measurement is performed on
sequences of $m$ molecules, each of which is taken from an ensemble $\{p_i,\rho_i\}$.
When $m$ tends to infinity $2^{m(S(\rho)-\sum_i p_i S(\rho_i))}$ typical sequences (the
sequences in which $\rho_i$ appears about $p_i m$ times) become mutually orthogonal and
can be distinguished by ``square-root" or ``pretty good" measurements
\cite{hausladen96_holevo98}. This situation is thus essentially classical, hence, $\Delta
S\rightarrow 0$ and the Holevo bound will be saturated.

An interesting implication of our result is the relationship between the second law and
the erasure principle. It has been shown that the form of the Holevo bound can be obtained 
from the erasure principle \cite{martin9901}. Together with this, our result
suggests that the erasure principle and the second law, which are commonly believed to be
equivalent, do not necessarily give the same result in the quantum regime.

As mentioned earlier in this paper, Land\'{e} has claimed that the superposition
principle in quantum mechanics can be derived by purely thermodynamical arguments that
are similar to our consideration \cite{lande52}. If this conclusion was correct even in
its spirit and if we could really derive quantum mechanics from thermodynamics, then we
should also be able to confirm the Holevo bound exactly. But, as shown here this is not
the case. This should, however, not necessarily be perceived as a failure of
thermodynamics. It is not unlikely that quantum theory will be superseded by a higher
level generalization of which it is a special case, just like classical mechanics is a
limiting case of quantum mechanics. Our paper shows that even if the amount of stored
information in the post-quantum theory can be greater than allowed quantum mechanically,
this can happen without violating the second law.

\begin{acknowledgments}
K.M. is a Boursier de l'ULB, and is supported by the European  Union through the project
RESQ IST-2001-37559, the Action de Recherche Concert\'{e}e de la Communaut\'{e}
Fran\c{c}aise de Belgique, and the IUAP program of the Belgian Federal Government under
grant V-18. {\v C}.B. has been supported by the European Commission, Marie Curie
Fellowship, Project No. 500764, and the Austrian Science Foundation (FWF) Project No.
F1506. V.V. acknowledges funding from EPSRC and the European Union.
\end{acknowledgments}

\end{document}